# AI as a Communication Facilitator: Shared Decision-Making Inspired Strategies for Bipolar Disorder Diagnosis and Treatment

AI as a Communication Facilitator for Bipolar Disorder Diagnosis and Treatment

AI Systems can be utilized to facilitate accurate and timely diagnoses and treatment decisions for Bipolar Disorder

Trisha Guttal

Cornell University, tkg32@cornell.edu

My HAI topic involves the use of AI communication facilitators to detect mood disorders such as bipolar disorder, a psychiatric condition in which patients experience drastic mood shifts. Due to the ill-defined nature of the disorder, it is difficult for even a psychiatrist alone to be confident with their diagnosis. Changes in mental and mood state are often highly subjective and difficult to pinpoint through short-term surveys and psychiatric consultations. For many patients, diagnosis and treatment based on trial-and-error is unavoidable. A timely and thorough diagnosis and treatment plan is associated with the need for an equal involvement of both the patient and the psychiatrist throughout the process. This conclusion is reached through a detailed assessment of current interventions for (i) the ill-defined nature of the disorder, and (ii) the trial-and-error requirement for medication and diagnosis. As a result, I propose the implementation of an AI communication facilitator that can aid in the appropriate diagnosis and treatment of bipolar disorder by embodying the shared decision-making model. I propose that the model can be broken down into specific critical decision points with considerations made for each party involved in the process, aligning with the service blueprint model. I conclude by emphasizing the importance of AI in bipolar disorder diagnosis and treatment due to its ability to embrace patient heterogeneity, and bridge the gap between mental healthcare and human-AI interaction.

CCS CONCEPTS • Human-AI Interaction • Human-Computer Interaction

**Additional Keywords and Phrases:** Service Blueprint Model, Shared Decision-Making, Medicines Optimization, Self-Management Plans for Medication Adherence, Patient Decision Aids, Clinical Decision Support

# 1 INTRODUCTION

Bipolar disorder is a psychiatric condition characterized by periods of depression and abnormally elevated moods. It is a leading cause of poor health worldwide, with a lifetime prevalence of 2.4% [32]. Even with treatment, about 37% of patients relapse into depression or mania within 1 year, and 60% within 2 years [18]. Due to the lack of a successful treatment combination for this disorder, diagnosis and treatment based on trial and error is unavoidable. As a result of the heavy rates of relapse, bipolar disorder is considered as a long-term illness with long-term effects. It is likely to become a permanent part of the patient's identity and status. Due to the ill-defined nature of the disorder, it is difficult for even a psychiatrist alone to be confident with their diagnosis. This issue is further complicated by the ranging definitions of the disorder, and its seemingly hidden "symptoms." Short-term surveys and psychiatric consultations provide a limited period for psychiatrists to evaluate and recognize early warning signs and mood swing stages for appropriate diagnosis and treatment of this complex disorder.

## 1.1 Vision Statement

A timely and thorough diagnosis is associated with the need for an equal involvement of both the patient and the psychiatrist throughout the process. On the one hand, it is important for patients to generate clear and expressive patient-reported data. On the other hand, it is also important that psychiatrists take the time to thoroughly investigate symptoms by continually checking in on the patient. Diagnosis and treatment decisions should happen early to prevent further issues in the patient's physical and mental state, but should also be accurate to eliminate chances of misdiagnosis. The accuracy of the diagnosis can be further validated by the AI system, through its facilitation of communication between both parties. Therefore, my vision statement is: AI communication facilitators can aid in the appropriate diagnosis and treatment of bipolar disorder by bridging gaps or misunderstandings between the patient and psychiatrist.

# 2 OVERVIEW OF BIPOLAR DISORDER: KEY PROBLEMS

In order to achieve the vision of accurate and timely diagnosis and treatment through effective patient-psychiatrist communication, it is important to explore two key problems that define the disorder today: (i) the ill-defined nature of the disorder, and (ii) the inevitable need to diagnose and prescribe patients based on trial-and-error methods.

## 2.1 The Ill-defined Nature of the Disorder

Bipolar disorder is often misdiagnosed for unipolar depression, an illness characterized by recurrent depressive episodes. In a chart review of 90 patients with affective disorders, this misdiagnosis occurred in 37% of patients [2]. As a result, it is evident that the distinction between bipolar disorder and unipolar depression is not clear-cut. Misdiagnosis of bipolar disorder has many consequences, including prescription of inappropriate drugs such as antidepressants, which might lead to more serious mood changes, and, ultimately, high health-care costs [30]. In a study of misdiagnosed patients, 78% of them developed more extreme conditions [23].

In the absence of definitive and objective biomarkers of physiological behaviors that are associated with conventionally defined bipolar disorder categories, appropriate and confident diagnosis and treatment is difficult, as emphasized by the prevalence of misdiagnosis [30]. Existing challenges are further compounded by the heterogeneity of bipolar disorder – no two patients have an identical experience [27]. The process of diagnosis is not standardized, and often involves trial and error of medications for treating depression versus bipolar disorder. It is only once the patient visits for multiple consultations and explains the effects of their selected medication that psychiatrists often have a better idea of which of the two illnesses they have.

## 2.2 Diagnosis and Medication based on Trial and Error is Unavoidable

As emphasized in the above section, The symptoms of the illness and the effects of medication can vary by patient and over time. The effects are often unpredictable. The treatment of bipolar disorder has a goal of acute stabilization – the point at which a patient with mania or depression arrives at a symptomatic recovery with euthymic (stable) mood [18].

Despite the existence of a clear goal for a bipolar patient's state through medication, the results of treatment are often suboptimal, and sometimes counter intuitive. 60 - 85% of bipolar patients under treatment experience a relapse in 4-5 years [20], residual symptoms, functional disability [4], and functional impairment issues [1]. Due to the common misdiagnosis of unipolar depression for bipolar disorder, antidepressant treatment is often prescribed. This medication may seem beneficial in the short-term, but research suggests an unfavorable risk/benefit relationship: adding antidepressants to a mood stabilizer provides minimal relief from bipolar disorder symptoms, while antidepressants alone carry risks of manic or hypomania relapses [19]. As a result, it is critical that patients and psychiatrists maintain communication throughout the diagnosis and treatment process, so possibilities of misdiagnosis are recognized early, and potential serious side-effects are limited.

## 3 EXISTING INTERVENTIONS FOR BIPOLAR DISORDER DIAGNOSIS AND TREATMENT

Medicine optimization is a framework targeting the above reasons for suboptimal treatment of bipolar disorder. It is coupled with patient-psychiatrist communication, and is a person-centered approach employed to ensure that patients obtain the "best possible outcomes from their medicines" so that they do less harm than good [41]. To employ medical optimization effectively, psychiatrists are recommended to be supported by appropriate tools including: (i) self-management plans for medication adherence, (ii) patient decision aids used in consultations involving medicines, and (iii) clinical decision support.

## 3.1 Self-management Plans for Medication Adherence

A self-management plan is a patient-centered approach for patients to manage their medicines [8]. This approach is especially applicable to patients with long-term conditions, such as bipolar disorder. The goal of this approach is to empower patients to be confident in managing their symptoms and illness themselves. This idea is based on the premise that patients who are actively involved in their health and well-being tend to manage their condition more effectively, and with a more positive outlook. This approach is often executed through electronic medication dosage reminder systems.

Bipolar disorder cannot be treated with self-management plans alone, however, as the success of this approach heavily depends upon the patient's determination. Electronic reminders can improve short-term medication adherence, but 61% of such tools show no evidence of significantly improving clinical outcomes [39, 22]. About half of the patients diagnosed with bipolar disorder become non-adherent during long-term treatment, a rate largely like other chronic illnesses, and one that has remained unchanged over the years [9]. Research has concluded that it is not electronic reminder systems that help maintain medication adherence, but rather the implementation of positive attitudes and collaborative treatment featuring decision-making with equal input from the patient and psychiatrist [9].

## 3.2 Patient Decision Aids used in Consultations Involving Medicines

A patient decision aid is a patient-centered approach designed to support patient decision-making by providing information about existing treatment or screening options and their associated outcomes [37]. This approach allows patients to feel as informed as their psychiatrists when navigating through multiple treatment options and medication reconciliation decisions. Patient decision aids can increase general knowledge, foster more patient involvement, improve patient-psychiatrist communication, and enhance the proportion of patients that select a treatment consistent with their values [37]. However, the long-term effects of decision-aids on patients resulted in no findings for a significant positive affect on overall health outcomes.

As a good practice for prescribing medication, it is important that patients utilize decision aids to involve their psychiatrist. Through effective discussions about the patient's values, goals and experiences with their illness, psychiatrist can gain a better understanding of whether their chosen treatment is an optimal next step. Maintaining this practice can improve medication adherence, allow both parties to gain a better understanding of next steps for diagnosis, and decrease existing symptoms [42].

### 3.3 Clinical Decision Support

A clinical decision support system (CDS) is a psychiatrist-centered component of an integrated clinical IT system providing support for clinical services to manage a patient's condition [43]. Most CDSs commonly feature alerts, reminders, order sets, drug-dose calculations, reminders for clinicians, or care summary dashboards with indicators on performance feedback or quality [5].

Although these systems are being used in many areas of medicine and have been associated with positive patient outcomes, their use in psychiatric cases specifically is limited [38]. This is primarily because of the constant advent of new medications for psychiatric illnesses. It is difficult for CDS technologies to catch up with these shifts – the mental healthcare industry continues to lag in the direct implementation of CDS capabilities. CDS technologies are also less frequently adopted because they do not tend to significantly reduce adverse effects of treatments [31] or mortality of bipolar disorder illness [14], and their interventions for mental health remain unclear [36]. Evidence to demonstrate the positive effects of CDSs on clinical and economic outcomes remains limited, resulting in rare adoption for bipolar disorder diagnosis and treatment [5].

CDS adoption is also limited due to various technical challenges. Current solutions are not sufficiently patient-specific, which is a heavy requirement for psychiatric illnesses compared to physiological illnesses due to their patient heterogeneity. Leveraging recent developments in machine learning can enhance CDS tools by generating new knowledge from gathered data to provide better patient specificity [25].

Collaborative Clinical Decision Support (CCDS) systems were created as a response to the limitations of traditional CDS approaches. Through this process, patients bring their experience, and psychiatrists bring their practical knowledge, therefore thinking beyond the traditional biophysical problem-solving techniques that CDS systems offer. The development of this system was based on the premise that systems are more likely to succeed if they involve both the patient and the physician, rather than being solely psychiatrist-centered [33]. The CCDS system therefore not only assists psychiatrists, but also empowers patients to play a significant role in their own care.

### 4 HOW FAR ARE WE FROM AI COMMUNICATION FACILITATORS?

One reason for these modest positive effects of bipolar patients might be that the technologies used for diagnosis and treatment exist in separate silos: one for patients (self-management plans for medication adherence, patient decision aids) and a different one for psychiatrists (CDSS). One thing we notice is that the effects of these technologies on diagnosis and treatment are more positive the more involvement the patient has in the process, and the more information is shared between the patient and psychiatrist. This conclusion, however, is based on limited evidence given the small number of systems that allow for direct collaboration between patients and psychiatrists for diagnosis and treatment.

In the long run, Artificially Intelligent (AI) systems can act as effective communication facilitators within this area, given their ability to build comprehensive theory and gather empirical data to support effective communication across the right points during the long-term diagnosis and treatment process. Additionally, AI systems can capture and tailor to the heterogeneity of the disorder - no patient's experience is the same [27]. However, reaching this point may take some time, given psychiatrists' slow adaptation to these technologies.

The mental healthcare industry has lagged in the direct implementation of machine learning algorithms and artificially intelligent models mostly because of the challenges of their integration with standard mental health care [3]. Within the area of communication between clinicians and patients specifically, AI has even more challenges to tackle. Communication is complex, with messages sent via verbal, para-verbal (e.g. voice tone) and non-verbal (e.g. eye-gaze, expressions) channels. Multiple participants are involved in the interaction, each with differing expectations, goals, roles, capabilities, and vulnerabilities [7]. Tasks within consultations can be diverse, including information gathering, education, decision-making, relationship-building and managing emotions. Therefore, building AI models around such nuanced interactions has been challenging [7].

Given the current low rates of AI adoption in mental healthcare and communication due to reasons emphasized above, it is recommended that AI be used as a "complement" to existing clinical practices and conversational settings rather than a "decision-maker" for diagnosis and treatment as a next step towards the vision [6]. This recommendation lays the foundation for considering the design of an AI system that facilitates conversation for arriving at the best healthcare outcomes for the specific bipolar patient, while also being supported by a Service Blueprint method, which is further discussed in the suggested solution space.

## 5 SHARED DECISION-MAKING

As discussed earlier in the paper, the key problems in diagnosing and treating bipolar disorder are (i) the lack of a clear definition of the disorder, and (ii) the continuous need to adopt a trial-and-error method for medication prescription. Both problems often overlap during the diagnosis and treatment process. As emphasized previously, the effects of siloed existing solutions for both problems tend to have low success rates. However, their effectiveness can be enhanced by the shared involvement of both the patient and psychiatrist through a clearly executed, long-term, step-by-step process.

In this section, I propose the employment of shared decision-making (SDM). It is defined as a process in which both the patient and psychiatrist work together to decide the best plan of care for the patient [15]. Shared decision-making provides the patient with a more active role in the diagnosis and treatment process, in which their values, goals and concerns are considered alongside the recommendations of the psychiatrist. This process is recommended as a starting point that Human-AI designers could take in attending to AI's potential role of facilitating productive communication between a patient and psychiatrist when diagnosing and treating bipolar disorder.

This idea is further emphasized by the long-term nature of the disorder: it may become a permanent part of the patient's identity and status. In this case, the patient-psychiatrist relationship can be considered long-term, with requirements of multiple consultations, monitoring, and adjustment of medications. The availability of various medications for treatment, sometimes prescribed in multiple permutations and combinations, requires that psychiatrists communicate closely with their patients to develop the optimal solution for diagnosis and treatment, a key principle of the shared decision-making model [41].

As bipolar disorder is a potentially life-threatening illness, it is critical that a patient receives timely diagnosis and treatment recommendations. As a result, there are certain critical treatment decision points in standard treatment pathways that may occur only once, and arise early on in the course of the illness which may have large consequences for the patient [34]. For instance, a bipolar patient may experience a new episode of acute depression, and must decide whether to adhere to existing medications, or perform medication reconciliation. Such a decision cannot be delayed without potentially serious implications for the health of the patient.

Shared decision-making is particularly important for addressing cases with critical decision points as, firstly, several treatment options for bipolar disorder exist, with different possible outcomes and substantial uncertainty of effectiveness. Secondly, there is no clear-cut right or wrong decision due to the ill-defined nature of the disorder. Thirdly, bipolar disorder is characterized by its heterogeneity - treatments will vary in their impact on each patient's physiological and psychological well-being [27]. This idea is further compounded by the patients' need for certainty that bipolar disorder is a treatable illness. It is common for patients to feel hopeless if medicines are ineffective, often leading to a lack of medication adherence which may dramatically worsen symptoms [26]. Therefore, the psychiatrist requires a strong understanding of the patient's goals and values first before suggesting a potential next step. Adopting the shared decision-making model gives psychiatrists a guideline for performing the right steps at the right time, therefore limiting patient harm.

### 5.1 Shared Decision-Making as a Potentially Short-term Solution

A strong, positive, and collaborative shared decision-making adoption is associated with "several positive bipolar disorder patient outcomes" including improved help seeking behavior [40], satisfaction in care [24], increased compliance with decisions [21], better medication adherence [17], reduced stigma [12], and reduced suicide risks [17]. However, none of the trials had investigated whether the conversations between patients and psychiatrists exhibited the "full range" of shared decision-making characteristics, which include defining the problem, outlining the options in detail, checking understanding, eliciting values, supporting deliberation, and reaching mutual agreement [11].

An even smaller number of trials looked at longer-term clinical outcomes. SDM is an ongoing relationship-building process that takes place between patient and psychiatrist - the model is not a tool or a one-off event, but rather a chain of conversations, arriving at a shared goal [11]. As many trials looked at only individual components of this inherently complex process of patient-psychiatrist interaction, it is difficult to assess SDM's effectiveness, especially for a long-term illness like bipolar disorder.

As a result, I urge that psychiatrists and patients adopt the shared decision-making model with the expectation of maintaining it as a long-term addition to the treatment process. Changing practice often involves the use of formal decision support tools to support SDM.

## 6 ADOPTING A HUMAN-AI SERVICE BLUEPRINT-BASED INTERACTION

In order to maintain the use of this model, specific decision points should be marked, so that they are readily accessible during clinical consultations. Additionally, it is recommended that "appropriate patient reported outcome measures (PROMs) be used in routine care as a feedback loop to check that patients are actively engaged and receive treatments that reflect their goals and preferences" as a step to comply with the long-term structure of the model [11]. Maintaining this feedback loop throughout the long-term for multiple patients with varying needs, however, can be extremely taxing. Some studies, such as one conducted on patient-physician interactions in France, found that physicians had difficulty maintaining a long-term commitment to the system, due to a lack of availability faced with the requirement of consistent and longitudinal communication with the patient [28]. Although this is a large issue with the implementation of SDM in bipolar disorder treatment and diagnosis, the arguments for adopting this approach, as discussed above, are much more compelling overall. The problem of challenges maintaining long-term commitments should be acknowledged and overcome if SDM is to be a dominant model in mental healthcare [11].

As a first step of overcoming these downfalls, one might be tempted to identify the specific critical decision points in the bipolar disorder trajectory for a patient, so that conversation between the patient and psychiatrist can occur at those specific time periods. This can be executed by employing a Service Blueprint - a diagram that visualizes the relationships between different service components, in this case, the patient and psychiatrist, and processes that are directly tied to touchpoints within the diagnosis and treatment process [16]. The service blueprint is an appropriate application to this problem as the aim is to increase the effectiveness of a long-term SDM process that features multiple parties. Mapping out these times can be beneficial as it allows the psychiatrist to keep track of their availability and maintain efficiency by tending to multiple bipolar patients at once.

Charles et al, the developer of SDM, however, makes a strong case that this SDM Service Blueprint approach has its limitations. He explains that there is no "single route" to shared decision-making; the heterogeneity of the bipolar patient experience must be acknowledged, and the SDM process cannot be reduced to a single, streamlined, step-by-step process [10]. If there was a way to maintain checkpoints that change and tailor to the specific patient's experience, however, we would be on track towards effective SDM conversations for diagnosis and treatment. This can be effectively executed by AI technologies due to their capability of developing tailor-made psychiatric care for the patient [27].

In order to move towards the effective implementation of long-term SDM practice for accurate and timely diagnosis of bipolar disorder, I propose the development of an AI communication facilitator system that is built upon a Service Blueprint framework. The system will utilize information about the patient (family history, number of relapses, types of treatments taken, duration of illness, etc.) to develop a longitudinal timeline featuring critical decision points in the diagnosis and treatment process for the specific patient. At these critical decision points, the AI system will play the role of a communication facilitator by stepping in and alerting both parties to engage in conversation about the ongoing treatment and diagnosis procedure. The outputs of this system will be equally accessible to both the patient and psychiatrist, therefore embodying the ethical justifications of the SDM model [35]. The AI system will continue to mold and change the long-term diagnosis timeline based on collective decisions made by both parties at the set critical decision points - no patient's AI-generated timeline will look the same.

Executing this AI solution requires a deeper consideration of the patient's and psychiatrist's experiences and requirements for the construction of a service blueprint. For instance, it is unclear whether the patient or psychiatrist should initiate the decision-making process, and whether either party has a preference for who should be taking these steps. Current studies on the preferences for and experiences of decision-making roles and involvement resulted in inconsistent findings, therefore signaling the need for more concrete research on these variables [17]. Given these initial implementation challenges, research in routine mental health services is required as a first step to arriving at the intended vision of an AI communication facilitator for effective diagnosis and treatment of bipolar disorder.

### 6.1 Limitations of Implementation

Although the AI system is intended to accommodate and tailor to all bipolar patients' experiences, it neglects the factor of patient involvement preferences. Older male patients were predicted to have stronger preferences for, and a greater likelihood of experiencing less involvement in decision-making in general [13]. When SDM stages were assessed separately, being Caucasian predicted preferences for less involvement and a higher reliance on psychiatrist decision-making, while higher education levels of patients predicted preferences for greater involvement [29].

However, recent studies on this topic [17] report a minority of patients who prefer a passive role in the diagnosis and treatment process (8.5% - 34.7%). Majority of patients preferred shared decision-making with their clinician (38.3% - 64.3%), or made the final decision alone (1.8% - 52.1%). In future iterations, the AI system can be polished further to accommodate for patient's specificity in involvement preferences. For instance, the number of key critical decision points that require both parties to be present can be reduced if the patient prefers making the final decision alone.

## 7 CONCLUSION

Artificial Intelligence will soon achieve such a level of technical capability that the shared decision-making model can be tailored to each specific long-term patient-psychiatrist relationship. Instead of adopting solutions that are siloed and either patient or psychiatrist-centered, using AI as a communication facilitator that adopts the Service Blueprint model will ensure that the patient and psychiatrist maintain timely and active communication throughout the diagnosis and treatment process. The execution of this facilitator system alongside the SDM model will reduce the impacts of the unclear definition of the disorder, whilst also supporting the trial-and-error procedure for treatment by giving the patient equal agency for their decisions.

As a first step to arriving at this ideal solution, it is recommended that more research be done on the details of the specific interactions involved in patient-psychiatrist interaction, so that an accurate starter service blueprint can be constructed.

I conclude by proposing that AI systems are designed to encourage diversity in patient and psychiatrist relationships. Their applications to areas of service blueprinting and shared decision-making suggest ways in which the bridge between mental healthcare and Human-AI interaction could be built.

## REFERENCES

[1] Lori L. Altshuler, Michael J. Gitlin, Jim Mintz, and Mark A. Frye. 2002. Subsyndromal Depression Is Associated With Functional Impairment in Patients With Bipolar Disorder. *J Clin Psychiatry* 63, 9 (September 2002), 0–0. Retrieved December 12, 2021 from https://www.psychiatrist.com/jcp/bipolar/subsyndromal-depression-is-associated-functional-impairment/

[2] Jules Angst and Giovanni Cassano. 2005. The mood spectrum: improving the diagnosis of bipolar disorder. *Bipolar Disorders* 7, s4 (2005), 4–12. DOI:https://doi.org/10.1111/j.1399-5618.2005.00210.x

[3] Luke Balcombe and Diego De Leo. 2021. Digital Mental Health Challenges and the Horizon Ahead for Solutions. *JMIR Mental Health* 8, 3 (March 2021), e26811. DOI:https://doi.org/10.2196/26811

[4] Christopher R. Bowie, Colin Depp, John A. McGrath, Paula Wolyniec, Brent T. Mausbach, Mary H. Thornquist, James Luke, Thomas L. Patterson, Philip D. Harvey, and Ann E. Pulver. 2010. Prediction of Real-World Functional Disability in Chronic Mental Disorders: A Comparison of Schizophrenia and Bipolar Disorder. *AJP* 167, 9 (September 2010), 1116–1124. DOI:https://doi.org/10.1176/appi.ajp.2010.09101406

[5] Tiffani J. Bright, Anthony Wong, Ravi Dhurjati, Erin Bristow, Lori Bastian, Remy R. Coeytaux, Gregory Samsa, Vic Hasselblad, John W. Williams, Michael D. Musty, Liz Wing, Amy S. Kendrick, Gillian D. Sanders, and David Lobach. 2012. Effect of Clinical Decision-Support Systems. *Ann Intern Med* 157, 1 (July 2012), 29–43. DOI:https://doi.org/10.7326/0003-4819-157-1-201207030-00450

[6] Matthias Brunn, Albert Diefenbacher, Philippe Courtet, and William Genieys. 2020. The Future is Knocking: How Artificial Intelligence Will Fundamentally Change Psychiatry. *Acad Psychiatry* 44, 4 (August 2020), 461–466. DOI:https://doi.org/10.1007/s40596-020-01243-8

[7] Phyllis Butow and Ehsan Hoque. 2020. Using artificial intelligence to analyse and teach communication in healthcare. *Breast* 50, (January 2020), 49–55. DOI:https://doi.org/10.1016/j.breast.2020.01.008

[8] NICE Medicines and Prescribing Centre (UK). 2015. *Self-management plans*. National Institute for Health and Care Excellence (UK). Retrieved December 12, 2021 from https://www.ncbi.nlm.nih.gov/books/NBK355902/

[9] Subho Chakrabarti. 2016. Treatment-adherence in bipolar disorder: A patient-centred approach. *World J Psychiatry* 6, 4 (December 2016), 399–409. DOI:https://doi.org/10.5498/wjp.v6.i4.399

[10] C. Charles, A. Gafni, and T. Whelan. 1997. Shared decision-making in the medical encounter: what does it mean? (or it takes at least two to tango). *Soc Sci Med* 44, 5 (March 1997), 681–692. DOI:https://doi.org/10.1016/s0277-9536(96)00221-3

[11] Angela Coulter. 2017. Shared decision making: everyone wants it, so why isn't it happening? *World Psychiatry* 16, 2 (June 2017), 117–118. DOI:https://doi.org/10.1002/wps.20407

[12] Niall Crumlish and Brendan D. Kelly. 2009. How psychiatrists think. *Advances in Psychiatric Treatment* 15, 1 (January 2009), 72–79. DOI:https://doi.org/10.1192/apt.bp.107.005298

[13] Carlos De las Cuevas, Wenceslao Peñate, and Luis de Rivera. 2014. Psychiatric patients' preferences and experiences in clinical decision-making: Examining concordance and correlates of patients' preferences. *Patient Education and Counseling* 96, 2 (August 2014), 222–228. DOI:https://doi.org/10.1016/j.pec.2014.05.009

[14] Øystein Eiring, Kari Nytrøen, Simone Kienlin, Soudabeh Khodambashi, and Magne Nylenna. 2017. The development and feasibility of a personal

health-optimization system for people with bipolar disorder. *BMC Med Inform Decis Mak* 17, 1 (July 2017), 102. DOI:https://doi.org/10.1186/s12911-017-0481-x

[15] Glyn Elwyn, Dominick Frosch, Richard Thomson, Natalie Joseph-Williams, Amy Lloyd, Paul Kinnersley, Emma Cording, Dave Tomson, Carole Dodd, Stephen Rollnick, Adrian Edwards, and Michael Barry. 2012. Shared Decision Making: A Model for Clinical Practice. *J Gen Intern Med* 27, 10 (October 2012), 1361–1367. DOI:https://doi.org/10.1007/s11606-012-2077-6

[16] World Leaders in Research-Based User Experience. Service Blueprints: Definition. *Nielsen Norman Group*. Retrieved December 12, 2021 from https://www.nngroup.com/articles/service-blueprints-definition/

[17] Alana Fisher, Vijaya Manicavasagar, Felicity Kiln, and Ilona Juraskova. 2016. Communication and decision-making in mental health: A systematic review focusing on Bipolar disorder. *Patient Education and Counseling* 99, 7 (July 2016), 1106–1120. DOI:https://doi.org/10.1016/j.pec.2016.02.011

[18] John R Geddes and David J Miklowitz. 2013. Treatment of bipolar disorder. *The Lancet* 381, 9878 (May 2013), 1672–1682. DOI:https://doi.org/10.1016/S0140-6736(13)60857-0

[19] S. N. Ghaemi, A. P. Wingo, M. A. Filkowski, and R. J. Baldessarini. 2008. Long-term antidepressant treatment in bipolar disorder: meta-analyses of benefits and risks. *Acta Psychiatrica Scandinavica* 118, 5 (2008), 347–356. DOI:https://doi.org/10.1111/j.1600-0447.2008.01257.x

[20] Michael J. Gitlin and David J. Miklowitz. 2017. The difficult lives of individuals with bipolar disorder: A review of functional outcomes and their implications for treatment. *Journal of Affective Disorders* 209, (February 2017), 147–154. DOI:https://doi.org/10.1016/j.jad.2016.11.021

[21] J. Hamann, S. Leucht, and W. Kissling. 2003. Shared decision making in psychiatry. *Acta Psychiatrica Scandinavica* 107, 6 (2003), 403–409. DOI:https://doi.org/10.1034/j.1600-0447.2003.00130.x

[22] Saee Hamine, Emily Gerth-Guyette, Dunia Faulx, Beverly B. Green, and Amy Sarah Ginsburg. 2015. Impact of mHealth Chronic Disease Management on Treatment Adherence and Patient Outcomes: A Systematic Review. *Journal of Medical Internet Research* 17, 2 (February 2015), e3951. DOI:https://doi.org/10.2196/jmir.3951

[23] Robert M. A. Hirschfeld and Lana A. Vornik. 2004. Recognition and diagnosis of bipolar disorder. *J Clin Psychiatry* 65 Suppl 15, (2004), 5–9.

[24] E. a. G. Joosten, C. A. J. de Jong, G. H. de Weert-van Oene, T. Sensky, and C. P. F. van der Staak. 2009. Shared Decision-Making Reduces Drug Use and Psychiatric Severity in Substance-Dependent Patients. *PPS* 78, 4 (2009), 245–253. DOI:https://doi.org/10.1159/000219524

[25] Gal Levy-Fix, Gilad J. Kuperman, and Noémie Elhadad. 2019. Machine Learning and Visualization in Clinical Decision Support: Current State and Future Directions. *arXiv:1906.02664 [cs, stat]* (June 2019). Retrieved December 12, 2021 from http://arxiv.org/abs/1906.02664

[26] Lydia Lewis. 2005. Patient perspectives on the diagnosis, treatment, and management of bipolar disorder. *Bipolar Disorders* 7, s1 (2005), 33–37. DOI:https://doi.org/10.1111/j.1399-5618.2005.00192.x

[27] Diego Librenza-Garcia, Bruno Jaskulski Kotzian, Jessica Yang, Benson Mwangi, Bo Cao, Luiza Nunes Pereira Lima, Mariane Bagatin Bermudez, Manuela Vianna Boeira, Flávio Kapczinski, and Ives Cavalcante Passos. 2017. The impact of machine learning techniques in the study of bipolar disorder: A systematic review. *Neuroscience & Biobehavioral Reviews* 80, (September 2017), 538–554. DOI:https://doi.org/10.1016/j.neubiorev.2017.07.004

[28] Nora Moumjid, Alain Brémond, Hervé Mignotte, Christelle Faure, Anne Meunier, and Marie-Odile Carrère. 2007. Shared decision-making in the physician-patient encounter in France: a general overview. *Zeitschrift für ärztliche Fortbildung und Qualität im Gesundheitswesen - German Journal for Quality in Health Care* 101, 4 (May 2007), 223–228. DOI:https://doi.org/10.1016/j.zgesun.2007.02.042

[29] Stephanie G. Park, Marisa Derman, Lisa B. Dixon, Clayton H. Brown, Elizabeth A. Klingaman, Li Juan Fang, Deborah R. Medoff, and Julie Kreyenbuhl. 2014. Factors Associated With Shared Decision–Making Preferences Among Veterans With Serious Mental Illness. *PS* 65, 12 (December 2014), 1409–1413. DOI:https://doi.org/10.1176/appi.ps.201400131

[30] Mary L Phillips and David J Kupfer. 2013. Bipolar disorder diagnosis: challenges and future directions. *The Lancet* 381, 9878 (May 2013), 1663–1671. DOI:https://doi.org/10.1016/S0140-6736(13)60989-7

[31] Sumant R. Ranji, Stephanie Rennke, and Robert M. Wachter. 2014. Computerised provider order entry combined with clinical decision support systems to improve medication safety: a narrative review. *BMJ Qual Saf* 23, 9 (September 2014), 773–780. DOI:https://doi.org/10.1136/bmjqs-2013-002165

[32] E. J. Regeer, M. Ten Have, M. L. Rosso, L. Hakkaart-van Roijen, W. Vollebergh, and W. A. Nolen. 2004. Prevalence of bipolar disorder in the general population: a Reappraisal Study of the Netherlands Mental Health Survey and Incidence Study. *Acta Psychiatrica Scandinavica* 110, 5 (2004), 374–382. DOI:https://doi.org/10.1111/j.1600-0447.2004.00363.x

[33] Pavel S. Roshanov, Natasha Fernandes, Jeff M. Wilczynski, Brian J. Hemens, John J. You, Steven M. Handler, Robby Nieuwlaat, Nathan M. Souza, Joseph Beyene, Harriette G. C. Van Spall, Amit X. Garg, and R. Brian Haynes. 2013. Features of effective computerised clinical decision support systems: meta-regression of 162 randomised trials. *BMJ* 346, (February 2013), f657. DOI:https://doi.org/10.1136/bmj.f657

[34] G. S. Sachs. 2004. Strategies for improving treatment of bipolar disorder: integration of measurement and management. *Acta Psychiatrica Scandinavica* 110, s422 (2004), 7–17. DOI:https://doi.org/10.1111/j.1600-0447.2004.00409.x

[35] Mike Slade. 2017. Implementing shared decision making in routine mental health care. *World Psychiatry* 16, 2 (2017), 146–153. DOI:https://doi.org/10.1002/wps.20412

[36] Nathan M. Souza, Rolf J. Sebaldt, Jean A. Mackay, Jeanette C. Prorok, Lorraine Weise-Kelly, Tamara Navarro, Nancy L. Wilczynski, R. Brian Haynes, and the CCDSS Systematic Review Team. 2011. Computerized clinical decision support systems for primary preventive care: A decision-maker-researcher partnership systematic review of effects on process of care and patient outcomes. *Implementation Science* 6, 1 (August 2011), 87. DOI:https://doi.org/10.1186/1748-5908-6-87

[37] Dawn Stacey, France Légaré, Nananda F Col, Carol L Bennett, Michael J Barry, Karen B Eden, Margaret Holmes-Rovner, Hilary Llewellyn-Thomas, Anne Lyddiatt, Richard Thomson, Lyndal Trevena, and Julie HC Wu. 2014. Decision aids for people facing health treatment or screening decisions. In *Cochrane Database of Systematic Reviews*, The Cochrane Collaboration (ed.). John Wiley & Sons, Ltd, Chichester, UK, CD001431.pub4.

DOI:https://doi.org/10.1002/14651858.CD001431.pub4

[38] Madhukar H. Trivedi, Janet K. Kern, Bruce D. Grannemann, Kenneth Z. Altshuler, and Prabha Sunderajan. 2004. A Computerized Clinical Decision Support System as a Means of Implementing Depression Guidelines. *PS* 55, 8 (August 2004), 879–885. DOI:https://doi.org/10.1176/appi.ps.55.8.879

[39] Marcia Vervloet, Annemiek J Linn, Julia C M van Weert, Dinny H de Bakker, Marcel L Bouvy, and Liset van Dijk. 2012. The effectiveness of interventions using electronic reminders to improve adherence to chronic medication: a systematic review of the literature. *J Am Med Inform Assoc* 19, 5 (September 2012), 696–704. DOI:https://doi.org/10.1136/amiajnl-2011-000748

[40] Penny Wakefield Sue Read Wilson Firth Jam. 1998. Clients' perceptions of outcome following contact with a community mental health team. *Journal of Mental Health* 7, 4 (January 1998), 375–384. DOI:https://doi.org/10.1080/09638239817978

[41] Introduction | Medicines optimisation: the safe and effective use of medicines to enable the best possible outcomes | Guidance | NICE. Retrieved December 12, 2021 from https://www.nice.org.uk/guidance/ng5/chapter/Introduction

[42] Good practice in prescribing and managing medicines and devices. Retrieved December 12, 2021 from https://www.gmc-uk.org/ethical-guidance/ethical-guidance-for-doctors/good-practice-in-prescribing-and-managing-medicines-and-devices

[43] 1 Recommendations | Medicines optimisation: the safe and effective use of medicines to enable the best possible outcomes | Guidance | NICE. Retrieved December 12, 2021 from https://www.nice.org.uk/guidance/ng5/chapter/1-Recommendations#clinical-decision-support